\documentstyle[preprint,prl,aps]{revtex}
\def\And{{\rm and\ }}

\newif\ifboo \boofalse

\def\Review#1{\boofalse{\it #1},}
\def\Name#1{{\sc #1},}
\def\Vol#1{\ifboo Vol. {\bf #1}\else{\bf #1}\fi}
\def\Year#1{\ifboo #1\else(#1)\fi}
\def\Book#1{\bootrue{\it #1},}
\def\Page#1{\ifboo {\rm p. #1}\else{\rm #1}\fi}
\input epsf
\begin{document}

\draft
\title{Depletion forces between two spheres in a rod solution.}
\author{K. Yaman$^*$, C. Jeppesen$^{**}$, C. M. Marques$^\dagger$}
\address{$^*$Department of Physics, U.C.S.B., CA 93106--9530, U.S.A.}
\address{$^{**}$Materials Research Laboratory,	U.C.S.B.,	CA 93106, U.S.A.}
\address{$^\dagger$C.N.R.S.-Rh\^one-Poulenc, Complex Fluids Laboratory UMR166\\
Cranbury, NJ 08512--7500, USA}
\date{\today}
\maketitle

\begin{abstract}
We study the  depletion interaction
between spherical particles of radius $R$ immersed in a dilute solution of 
rigid rods of length $L$. The computed interaction potential  
is, within numerical accuracy,  exact for any value of
$L/R$. In particular we find that for $L \sim R$, the depth of the
depletion well is {\sl smaller} than the prediction of the
Derjaguin approximation. 
Our results bring new light into the discussion on the lack of phase
separation in colloidal mixtures of spheres and rods. 
\end{abstract}

\vskip 2truecm

Mixtures of colloids are abundant in industry as paints, glues,
lubricants and other materials~\cite{russel}. They are also
present  in the preparation of  foods and drugs, and in the
biological realm: many living organisms or components of living organisms
are colloidal suspensions of a variety  of sizes and shapes. 
A pervasive issue 
in the colloidal domain is the stability of colloidal
suspensions. Stability is often necessary for practical purposes as in
paint formulation for instance, but certain applications like water
treatment or mineral recovery might instead require aggregation or
flocculation to occur. The determination of the  stability criteria or the
study of the flocculation kinetics  can be achieved with reasonably good
accuracy once the interparticle interaction potentials are 
known~\cite{klein}. 
In a
system of a {\em pure} colloidal species  the two main interactions are the
van-der-Waals attraction and the hard-core repulsion. Such a
system is intrinsically unstable, the van-der-Waals 
attractive component of the
potential always leading to flocculation. 
In practice the stabilization of the
suspension  is enforced by using the screened electrostatic repulsion in 
aqueous solutions or  steric polymer layers in organic solvents. To a good
approximation the stabilized suspension can then be regarded as hard-core
particles with no attractive potential component.
One way of treating the stability issue in a  colloidal {\em mixture} 
of two components is by considering
the effective potential that the second species induces between two
particles of the first species. 

\begin{figure} [t]
\epsfxsize=4.65in
\epsfysize=2.55in
\centerline{\epsfbox{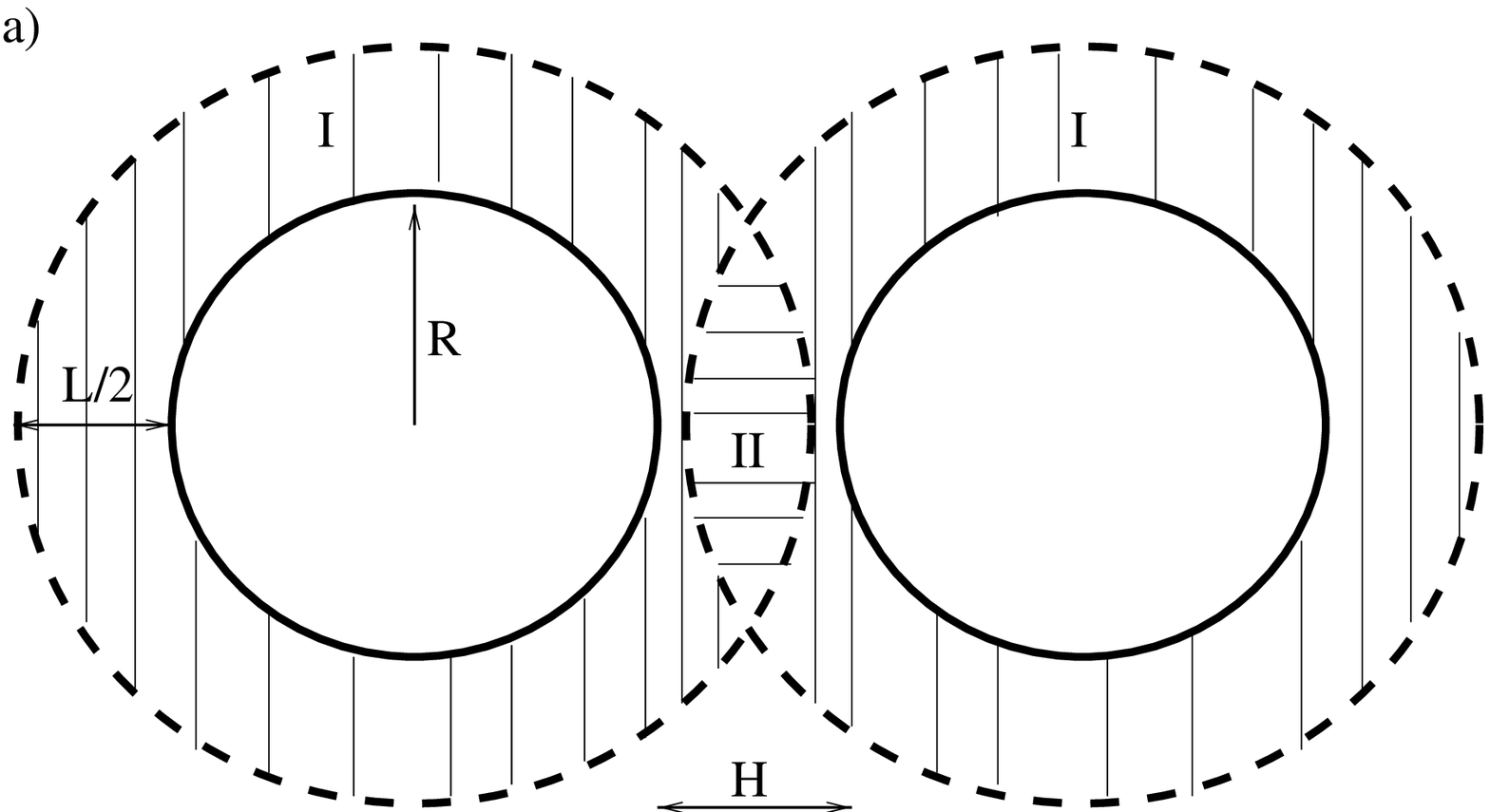}}
\epsfxsize=4.65in
\epsfysize=2.55in
\vspace{1in}
\centerline{\epsfbox{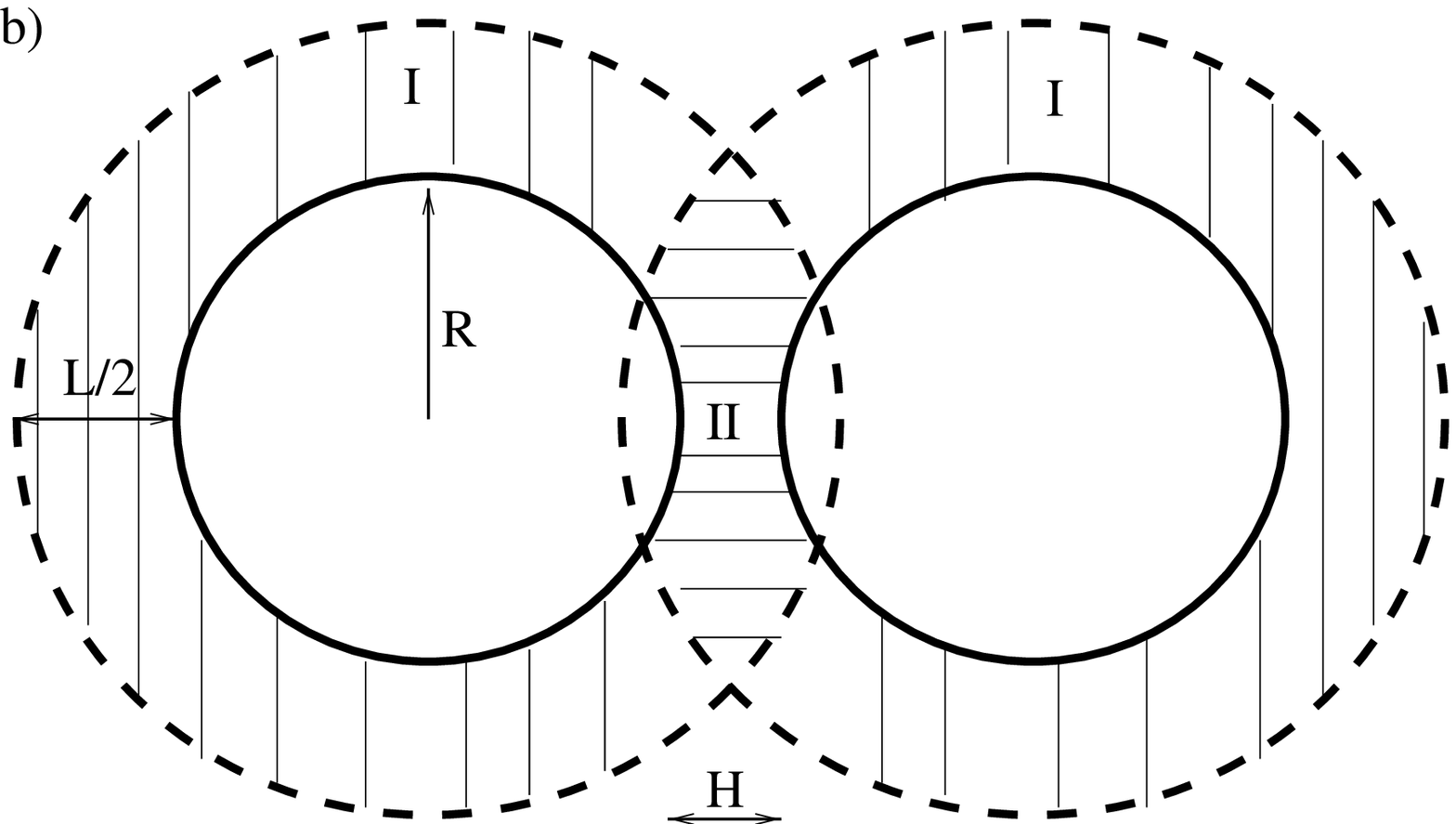}}
\vfil
\caption{The depletion region: {\it a)} for $H > L/2$, 
{\it b)} for $H <L/2$.}
\label{figregion}
\end{figure}

A well studied case~\cite{asakura2,vrij} is the attractive
potential that a solution of hard-core spheres of radius $r_0$ induce
between two spheres of radius $R$, (see fig.~\ref{figregion}):
\begin{equation}%%%%%%%%%%%%%%%%%%%%%%%%%%%%%%%%%%%%%%%%%%%%%
U_s(H) = - k_B T \frac{3}{8} \, \phi \, \frac{R}{r_0} \left( 2 -
\frac{H}{r_0}\right)^2
\left( 1 + \frac{2}{3}\frac{r_0}{R}+\frac{1}{6} \frac{H}{R}\right) 
\mbox{; for } \, H \leq 2 r_0
\label{vrij}
\end{equation}%%%%%%%%%%%%%%%%%%%%%%%%%%%%%%%%%%%%%%%%%%%%%%%
where $H$ is the distance between the two large spheres and $\phi$ the
volume fraction of the small spheres. Of course, $U_s$ is zero if 
$H \geq 2 r_0$; by definition of the depletion potential the depletion
at $H = 2 r_0$ is the zero-point. 
We will only give the potential in the regions where 
it is non-zero from now on. 
The depletion attraction has
been known since the pioneering work~\cite{asakura} 
of Asakura and Oosawa who
calculated the interaction energy 
between two flat plates immersed in a hard-sphere
solution:
\begin{equation}%%%%%%%%%%%%%%%%%%%%%%%%%%%%%%%%%%%%%%%%%%%%%
u_f(H) = - \frac{k_B T}{r_0^2} \frac{3}{4 \pi} \phi 
\left(2 - \frac{H}{r_0} \right)
\label{flatspheres}
\end{equation}%%%%%%%%%%%%%%%%%%%%%%%%%%%%%%%%%%%%%%%%%%%%%%%
The depletion potential has a purely entropic origin: 
the exclusion of the small
particles from the gap creates a pressure
deficit and hence an effective attraction between the plates.
When the exact analytical form of the interaction is only known for flat
plates, one can still compute the interaction potential 
between spheres much larger than the range of the potential by
employing the so-called Derjaguin approximation~\cite{israelachvili}.
For instance, the asymptotic behaviour (as $R/(2 r_0) \gg 1$) 
of eq.~(\ref{vrij}) can be obtained from eq.~(\ref{flatspheres}):
\begin{equation}%%%%%%%%%%%%%%%%%%%%%%%%%%%%%%%%%%%%%%%%%%%%%
U_{Der.}(H) = - \pi R \int_H^\infty dH' u_f(H') = 
- k_B T \frac{3}{8} \, \phi \, \frac{R}{r_0} \left( 2 -
\frac{H}{r_0}\right)^2
\label{derjaguin}
\end{equation}%%%%%%%%%%%%%%%%%%%%%%%%%%%%%%%%%%%%%%%%%%%%%%%
Note that the Derjaguin approximation underestimates the attraction in
this case.
Indirect experimental
evidence for the depletion effect were since long ago 
provided by the flocculation observed
in colloids and emulsions, but more direct observations of the
attraction have been only recently 
performed by force measurements~\cite{richetti} with surface force
apparatus or microscopy-techniques~\cite{dinsmore}. 

A much less-studied case is the depletion induced by hard-core rods, a
surprising fact considering that rod-shaped particles in the
colloidal range are present in a large variety of mineral and
organic systems~\cite{buining};
they also exist in the biological realm
where examples  range from  TMV-like virus  to fibrils of amyloid
$\beta$-protein, the molecular agent at the origin of the Alzheimer disease.
The depletion interaction between two flat surfaces immersed in a
dilute rod solution was studied in~\cite{asakura}:
\begin{equation}%%%%%%%%%%%%%%%%%%%%%%%%%%%%%%%%%%%%%%%%%%%%%
u_f(H) = - \frac{k_B T}{L D} \, 4.2 \, 
\frac{\phi}{\phi^*} \frac{1}{2} 
\left(1 - \frac{H}{L}\right)^2
\label{flatrods}
\end{equation}%%%%%%%%%%%%%%%%%%%%%%%%%%%%%%%%%%%%%%%%%%%%%%%
with $D$ the diameter of the rod, and $\phi^*$ the Onsager volume
fraction, below which the solution is isotropic. For spheres in a rod
solution the potential can be written in general as:
\begin{equation}%%%%%%%%%%%%%%%%%%%%%%%%%%%%%%%%%%%%%%%%%%%%%
U_s(H) = k_B T \, 4.2 \, 
\frac{\phi}{\phi^*}\,\frac{R}{D}\, K_1(H/L)
\label{depletion}
\end{equation}%%%%%%%%%%%%%%%%%%%%%%%%%%%%%%%%%%%%%%%%%%%%%%%
but no explicit form of $K_1$ is available except in the Derjaguin
approximation:
$K_1^{Der.}(x) = -(\pi/6)\, \left(1 - x \right)^3$,
which is expected to apply only 
for $L/R \ll 1$. Since this potential has quite a large contact value
compared to $k_B T$, one would expect on this basis to observe
flocculation also in rod/sphere mixtures. Surprisingly, such a
phase separation has not yet been clearly identified experimentally. 
A possible explanation for this lack of experimental evidence has
recently been proposed~\cite{mao}, based on repulsive contributions
from the surface modification of the rod-rod excluded volume
interactions. However such contributions are second order in the 
rod-concentration, and thus small below the Onsager
concentration~\cite{onsager},
$\rho_b^* = 4.2 / (L^2 D)$. The authors of reference~\cite{mao} also
note that 
most systems of interest involve rods that are of
comparable size to the spherical particles; in particular the
experiments~\cite{tracy} cited in~\cite{mao} have $1.16 < L/R < 4.34$,
a range where the use of the Derjaguin approximation to compute the
depletion potential in questionable.

In this letter we present {\em exact} results for the depletion
potential to first order in rod density. Our results are numerical in
nature, but exact, i.e.  the  inaccuracy of  our calculation is only
limited by  the numerical nature of the integration performed. It is
for our case  reducible to levels at which the error bars on the
figures given below are not visible.  The method of our
calculation is based on the work on surface tension of
objects immersed in rigid rod solutions that we published
elsewhere\cite{yaman2,yaman3}. 
We now briefly discuss this method and present our results.

We consider two spheres of radius $R$ in a solution of rods of 
length $L$, and thickness $D$, with $L/D \gg 1$. We work to first
order in the density of the rods, hence we really consider one rod and
two spheres.
A priori this is only valid when the rod-rod
interactions can be ignored: for rod concentrations below the
threshold $\rho_b^*$.

\begin{figure} [t]
\epsfxsize=3.8in
\epsfysize=4.7in
\centerline{\epsfbox{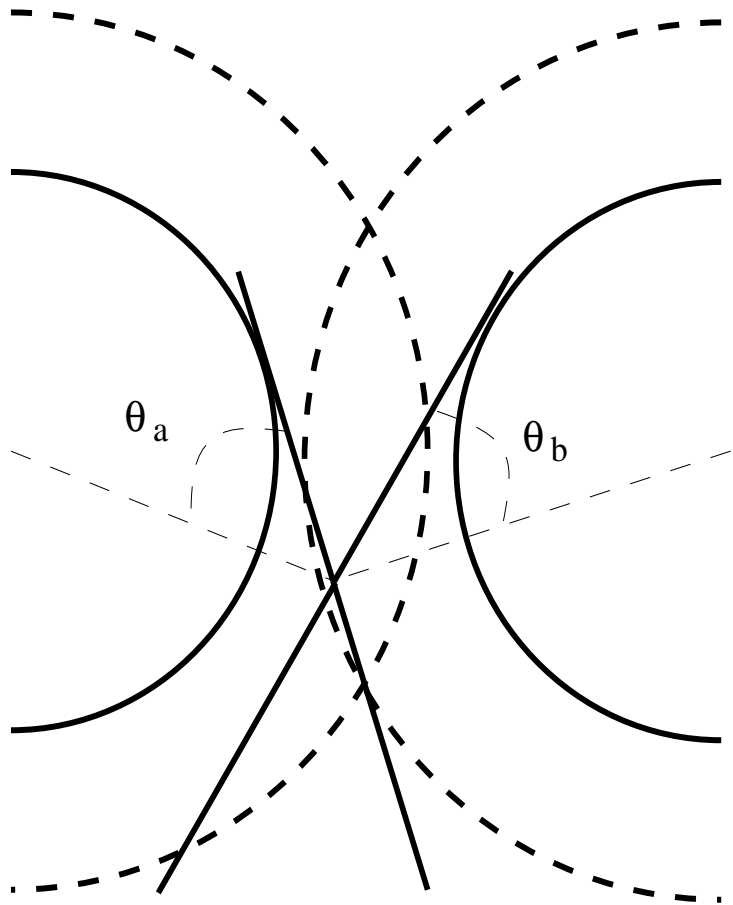}}
\vfil
\caption{The basic construction of the numerical integration method.}
\label{figangles}
\end{figure}

The interaction potential is given~\cite{yaman3} 
by the differences in the grand
potentials of the spheres at infinite separation and inside the
range of the potential, $L$. 
These grand potentials are integrals over 
the phase space available to a rigid rod: 
If the rod's center of mass is at a perpendicular 
distance $z$ greater than $L/2$ to the surface of a sphere, 
then the rod is free to rotate, otherwise
its rotational degrees of freedom are limited. The ``depletion
region'' of the rod is shown in fig.~\ref{figregion}. This region can
be further broken into two: In the first one (I) 
the rod interacts only with one sphere, whereas in the second one (II) 
interactions with both spheres are possible. 
Our previous work~\cite{yaman2} provides exact expressions for the 
contribution of
region I to the free energy for all values of $(L/R)$.
Region II needs to be further broken down
to three sub-regions where the functional form of the phase space
allowed to the rod is different. The integration over the angular
degrees of freedom can still be performed analytically, but
integration over the possible center of mass positions is complicated
by the lack of analytical expressions for the boundaries between the
sub-regions where different functional forms apply. We therefore use
the following procedure:
once the position of the center of mass is given,
we compute two angles, $\theta_a$ and $\theta_b$, (see
fig.~\ref{figangles}), from its coordinates, 
and then numerically decide
what functional form to use. Since we have exact expressions for the
allowed phase spaces, this procedure reduces the problem to a
two-dimensional numerical integration from a four-dimensional one,
which can be performed rapidly on a personal computer.

\begin{figure} [t]
\epsfxsize=5.0in
\epsfysize=4.0in
\centerline{\epsfbox{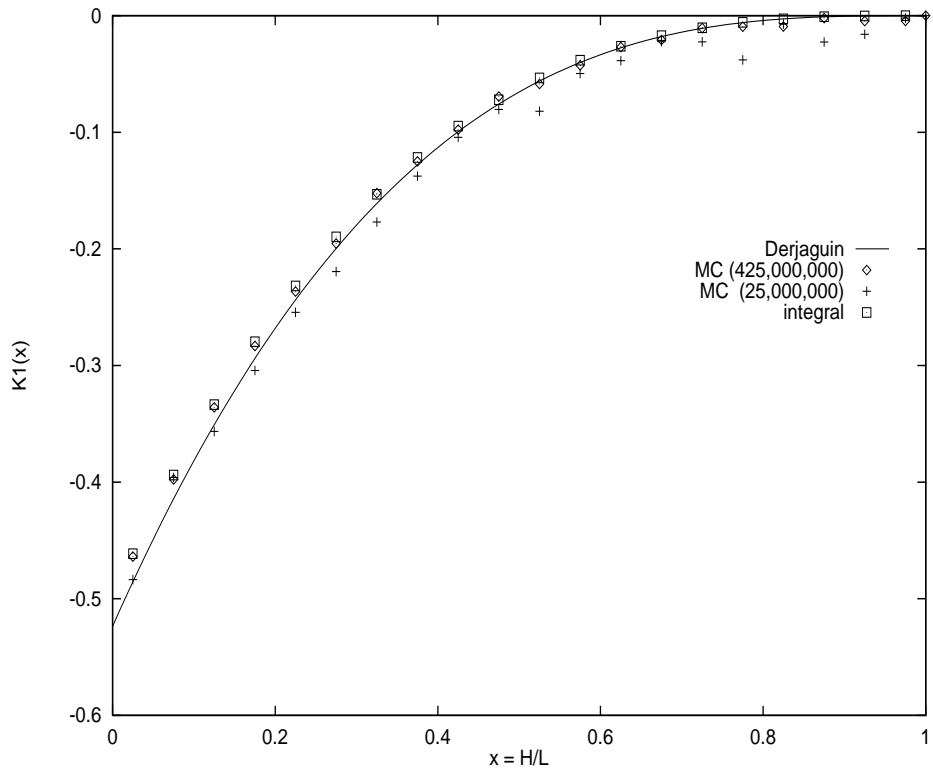}}
\vspace{5cm}
\vfil
\caption{Monte-Carlo compared to the numerical integral
for $L/R = 0.1$.}
\label{figplot2}
\end{figure}

One can, of course, do the calculation using Monte-Carlo integration, 
and we have checked our results using this technique. 
When region II is not
too small as compared to the total depletion region, Monte-Carlo
yields reasonable accuracy, moderately fast. The results for these
small $H$ values are in perfect agreement, (see fig.~\ref{figplot2}),
with the exact method, for
values of $(L/R)$ anywhere from $0.1$ to $15.0$. 
When $H$ gets larger, though, Monte-Carlo results 
fluctuate considerably; longer machine time is needed to reduce these
fluctuations.

\begin{figure} [t]
\epsfxsize=5.0in
\epsfysize=4.0in
\centerline{\epsfbox{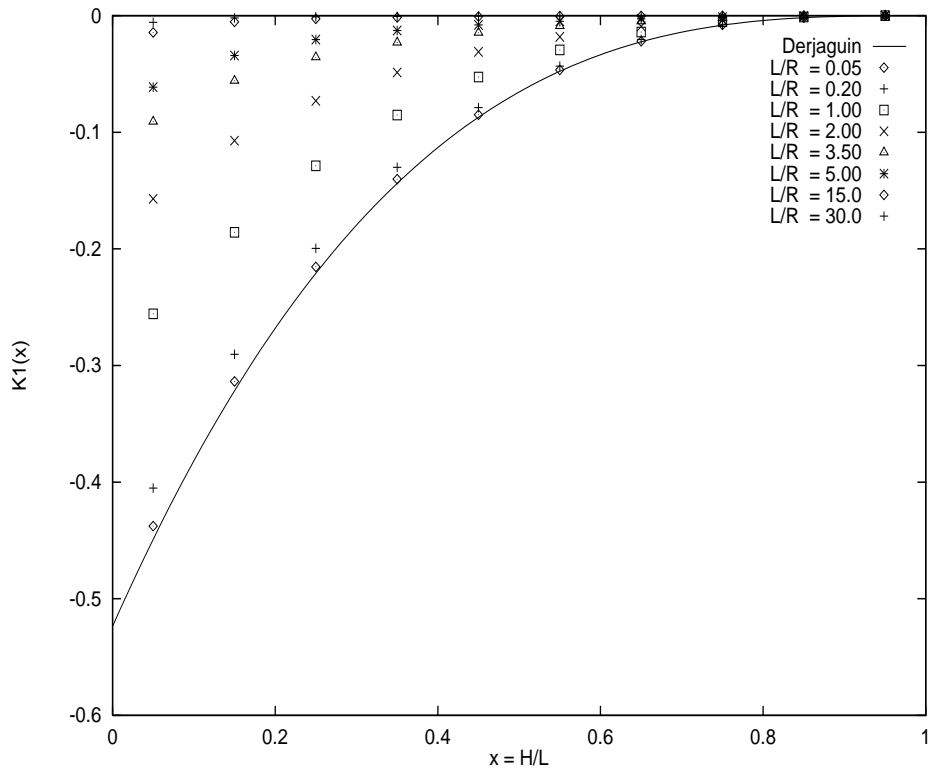}}
\vspace{5cm}
\vfil
\caption{The function $K_1(x)$ for several values of $L/R$.}
\label{figplot}
\end{figure}

\begin{figure} [t]
\epsfxsize=5.0in
\centerline{\epsfbox{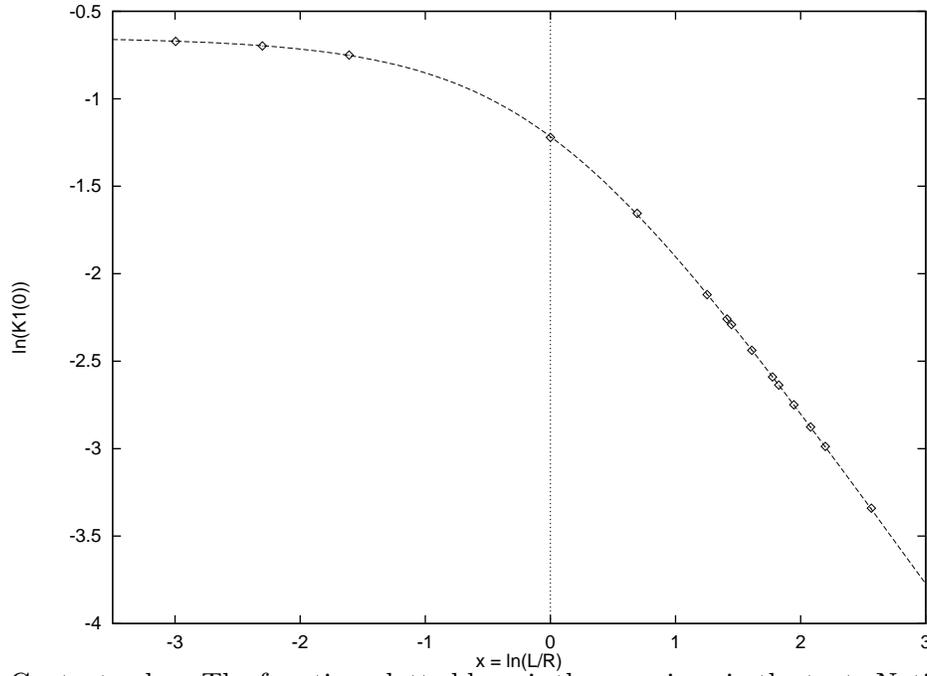}}
\caption{Contact values.The function plotted here is the one given in
the text. Notice the drastic change in the functional form of the data
when $L$ crosses $2 R$.}
\label{figcontact}
\end{figure}

Representative results for the function $K_1(x)$ defined in
eq.~(\ref{depletion}) are shown in fig.~\ref{figplot}. 
The Derjaguin function, $K_1^{Der.}(x)$ is also shown for comparison. 
As can be seen from this plot, Derjaguin is indeed an
excellent approximation to the depletion potential when 
$(L/R) \ll 1$. What is noteworthy is the {\em large} and systematic
deviation from this approximation when $(L/R)$ grows. This
deviation is about $50 \%$ when $(L/R) = 1$, and becomes more
pronounced as $(L/R)$ increases. 
This can be understood in the light of
general results we presented~\cite{yaman3} recently. These results
relate the second derivative of the free energy of a rod system to
the measure of configurations where the rod can wedge between two
points on the surface. Clearly, in our case for a fixed $H$, 
this measure goes to zero as the rod gets longer. 
This implies that
the free energy becomes a linear function of $(L/R)$,
and the function $K_1$, being related to the free energy by a factor
of $(L/R)^2$, vanishes as $R/L$.
Figure~\ref{figcontact} shows the behaviour of the contact value,
$K_1(0)$, as a function of the rod/sphere ratio. We fitted the data of
fifteen points to a function with three parameters. The resulting fit
is: 
\begin{equation}
K_1(0) = - \frac{\pi}{6} \, 
\frac{1 + 0.8762 \,(L/R)}{1 + 1.33198\, (L/R) + 0.98225\, (L/R)^2}
\label{contact}
\end{equation}
as can be seen in fig.~\ref{figcontact} the fit is very good
indeed for the range $0.05 < L/R < 13$.

The lack of phase separation in the experiments~\cite{tracy} 
is less surprising in the light of our results. Previous
expectations based on Derjaguin approximation estimate
the value of the attraction minimum at $10-20 k_B T$. Our results show
that the real value is much lower.
This reduction amounts to a factor
of three to five for these experiments and reduces the
depth of the depletion potential to a few $k_B T$.
Note also that other factors like chain flexibility
are likely to reduce the strength of the attraction further.  

We thank Phil Pincus for useful discussions.
KY acknowledges support from the National Science Foundation 
(NSF) awards DMR96-32716, and DMR8-442490-22213-3, 
CJ from NSF Grant CDA96-01954.

\newpage

%\noindent FIGURE CAPTIONS:\\

%\noindent The depletion region: {\it a)} for $H > L/2$, 
%{\it b)} for $H <L/2$.
%\medskip

%\noindent The basic construction of the numerical integration method.
%\medskip

%\noindent Monte-Carlo compared to the numerical integral
%for $L/R = 0.1$.
%\medskip

%\noindent The function $K_1(x)$ for several values of $L/R$.
%\medskip

%\noindent Contact values.The function plotted here is the one given in
%the text. Notice the drastic change in the functional form of the data
%when $L$ crosses $2 R$.

%\newpage

%%%%%%%%%%%% FIGURES %%%%%%%%%%%%%%%

\newpage

\newpage

\newpage

\newpage

\newpage

\end{document}